\documentclass[aps,twocolumn,showpacs,superscriptaddress]{revtex4-1}
\usepackage{amsmath,amssymb}
\usepackage{graphics,graphicx,color}
\usepackage{dcolumn,bm}
\usepackage{tikz}
\usepackage[colorlinks=true,citecolor=blue,urlcolor=blue]{hyperref}
\usepackage[mode=buildnew]{standalone}
\usepackage{tabularx}

\def\[{\left[}
\def\]{\right]}
\def\({\left(}
\def\){\right)}
\def\be{\begin{equation}}
\def\ee{\end{equation}}
\def\bea{\begin{eqnarray}}
\def\eea{\end{eqnarray}}

\begin{document}

\title{Shear Induced Orientational Ordering in Active Glass}

\author{Rituparno Mandal}
\affiliation{Institute for Theoretical Physics, Georg-August-Universit\"{a}t G\"{o}ttingen, 37077 G\"{o}ttingen, Germany.}
\author{Peter Sollich}
\affiliation{Institute for Theoretical Physics, Georg-August-Universit\"{a}t G\"{o}ttingen, 37077 G\"{o}ttingen, Germany.}
\affiliation{Department of Mathematics, King's College London, London WC2R 2LS, UK}

\begin{abstract}
Dense assemblies of self propelled particles, also known as active or living glasses are abundant around us, covering different length and time scales: from the cytoplasm to tissues, from bacterial bio-films to vehicular traffic jams, from Janus colloids to animal herds. Being structurally disordered as well as strongly out of equilibrium, these systems show fascinating dynamical and mechanical properties. Using extensive molecular dynamics simulation and a number of different dynamical and mechanical order parameters we differentiate three dynamical steady states in a sheared model active glassy system: (a) a disordered phase, (b) a propulsion-induced ordered phase, and (c) a shear-induced ordered phase. We supplement these observations with an analytical theory based on an effective single particle Fokker-Planck description to rationalise the existence of the novel shear-induced orientational ordering behaviour in our model active glassy system that has no explicit aligning interactions, {\textit{e.g.}}\ of Vicsek-type. This ordering phenomenon occurs in the large persistence time limit and is made possible only by the applied steady shear. Using a Fokker-Planck description we make testable predictions without any fit parameters for the joint distribution of single particle position and orientation. These predictions match well with the joint distribution measured from direct numerical simulation. Our results are of relevance for experiments exploring the rheological response of dense active colloids and jammed active granular matter systems.
\end{abstract}

\maketitle

Glassy or slow dynamics has been observed and thoroughly investigated in recent years in dense living or synthetic active matter systems across a range of scales, {\textit{e.g.}}\ dense assemblies of cells~\cite{angelini11}, crowded cellular cytoplasm~\cite{parry14}, glassy liquids formed by self-propelled Janus colloids~\cite{leomach19a,leomach19b} and jammed active granular solid~\cite{sriram14}; see Refs.~\cite{berthier19,janssen19} for recent reviews. Simulations and experiments have shown many non-trivial dynamical signatures, {\textit{e.g.}}\ active jamming~\cite{silke11}, shape dependent fluidization in a self-propelled Voronoi model~\cite{bi15,bi16}, glassy swirls in active dumbbells~\cite{mandal17}, intermittent dynamics with transient jamming~\cite{mandal20}, non-monotonic response in a glassy assembly of Janus colloidals~\cite{leomach19a,leomach19b}, along with the strong dynamical heterogeneity and slow density relaxation~\cite{berthier13,ni13,berthier14} that are typical to passive supercooled liquids. Although dynamical aspects and transport in such active glassy systems have been looked at quite extensively, the response of such out-of-equilibrium systems to mechanical perturbation, i.e.\ their rheology, remains largely unexplored. Recently Barrat et.\ al.~\cite{barrat17} have investigated the rheological response of a particular model dense active material that mimics the dynamics of tissue. In this model the number of particles is not conserved and particle death (apoptosis) and birth are modelled as a stochastic process.
\begin{figure}
\centering
\includegraphics[width=\columnwidth]{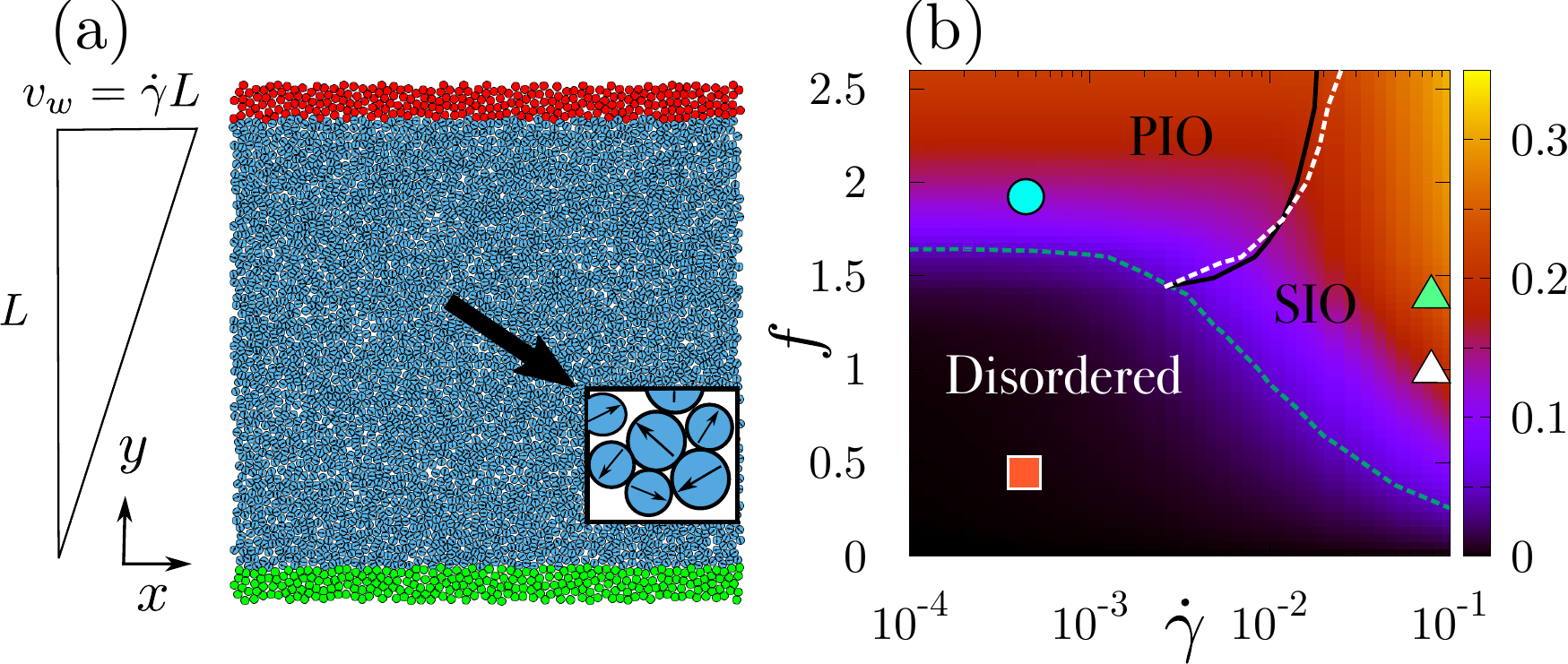}
\caption{(a) Schematic of the active glass under steady shear, showing the upper wall (red particles) that moves steadily with velocity $v_w$, the static lower wall (green particles) and the bulk (blue particles); arrows represent the active forcing direction on each particle. (b) Phase diagram constructed from numerical MD simulation of the sheared active glass, showing different dynamical phases: (i) {\em Disordered}: orientationally disordered phase, (ii) {\em PIO}: propulsion-induced ordered phase and (iii) {\em SIO}: shear-induced ordered phase. The dashed blue line marks the boundary between the orientationally ordered and disordered phases; the solid black and dotted white lines delineate the separation between the PIO and the SIO phase according to two complimentary criteria (see text for details).}
\label{fig:phased}
\end{figure}
The authors of~\cite{barrat17} observe a Newtonian crossover in the flow curves (stress versus strain rate) which is set by the cell death rate. In contrast to widely studied active particle classes 
({\textit{e.g.}}\ active Brownian particles (ABP)~\cite{marchetti12,takatori15,levis17,solon18} or active Ornstein-Uhlenbeck particles~\cite{szamel14,maggi15}), the particles in~\cite{barrat17} lack orientational degrees of freedom and active propulsion forces. The rheological response of active Brownian particles has also been explored~\cite{baskaran19} and an interesting velocity reversal phenomenon was observed near the boundary, though this study was limited to low-density suspensions of self propelled particles. Further studies exist of the shear response of active polar~\cite{luca10,cates19} systems, again in the dilute regime, as well as nematic systems~\cite{muhuri07,cates08} where liquid-crystalline  ordering phenomena appear. But the overall understanding of the response to steady shear of active glassy systems made of isotropic particles with orientational degrees of freedom remains an open question in the literature.

In this article we report a new type of orientational ordering in a glassy assembly of active Brownian particles (ABP). This orientational order appears without any mutual alignment interaction between the particles (of {\textit{e.g.}} Vicsek type~\cite{vicsek95} or arising from anisotropic particle shapes) and is facilitated by shear. Using different physical quantities in steady shear, specifically (a) non-affinity in the velocity profile, (b) shear stress and (c) an orientational order parameter, we can differentiate between a number of qualitatively distinct steady states. Even though shear might naively be thought of as injecting additional fluctuations into the system and thus suppressing ordering, we find that it can in fact help the system to order orientationally. The three phases we observe are: (i) a disordered phase, (ii) a propulsion-induced ordered (PIO) phase, and (iii) a shear-induced ordered (SIO) phase. To understand the most intriguing state, which is the shear-induced ordered phase, we use a Fokker-Planck equation as an effective description of the single particle dynamics. Using a Galerkin truncation we evaluate the steady state joint probability distribution $P(\theta,y)$ of particle position (in the shear gradient direction) and orientation in closed form. The theoretically predicted distributions $P(\theta,y)$, for which all parameters can be determined independently, are in very good qualitative agreement with our simulation results in the shear-induced ordered phase.

\section*{Model}

We study a model active glass~\cite{mandal20,mandalprl20} that can be viewed as essentially a passive glass former but with the dynamics of each particle driven by a self-propulsion force in addition to the usual force from interaction with its neighbours. For the underlying passive glass former we use the well-known Kob-Andersen model~\cite{kob95,bruning08}, which is a $65:35$ binary mixture of soft particles with non-additive Lennard-Jones interactions. For their dynamics we assume an active Brownian particle form~\cite{marchetti12,takatori15,levis17,solon18} (in terms of active forcing) with added steady shear:
\begin{equation}
m{\dot{\mathbf{v}}}_i=-\gamma (\mathbf{v}_i-\mathbf{v}_{F}(y_i)) + \mathbf{F}_{i} + f \mathbf{n}_i.
\label{eq_of_motion}
\end{equation}
Here $m$ is the mass of each particle and $\gamma$ the friction coefficient; we set both to $1$ in our study. We denote by $\mathbf{v}_i$ the velocity of the $i$-th particle and by $\mathbf{v}_{F}(y_i)$ the local affine flow velocity from the applied shear; viscous damping is taken as relative to this flow and can thus be viewed as resulting from an affinely sheared solvent. With an imposed shear rate $\dot{\gamma}$ the affine velocity field is $\mathbf{v}_{F}(y)=\dot{\gamma} y\hat{\bf{e}}_x$ if $y$ denotes the position coordinate along the shear gradient direction and $\hat{\bf{e}}_x$ is the unit vector in the shear ($x$-)direction. To implement the constant shear rate we move all the particles in an upper wall with a constant velocity $v_w$ while the particles in the corresponding lower wall are static. The upper wall velocity then sets the imposed shear rate through the relation $\dot{\gamma}={v_w}/{L}$ where $L$ is the distance between the upper and lower walls; see the illustration in Fig.~\ref{fig:phased}(a). We return below to the importance of implementing the steady shear with explicit walls.

Returning to the remaining terms in \eqref{eq_of_motion}, ${\bf{F}}_i$ is the total force on particle $i$ resulting from the Kob-Andersen LJ interactions~\cite{kob95,bruning08}. The active force has constant magnitude $f$ and direction $\mathbf{n}_i=(\cos\theta_i,\sin\theta_i)$ for the $i$-th particle. Each angle $\theta_i$ changes diffusively, $\dot\theta_i=(2/\tau_p)^{1/2}\eta_i(t)$ with $\eta_i(t)$ unit variance white noise; $\tau_p$ is thus the persistence time of the active forcing~\cite{mandal20,mandalprl20}.
\begin{figure}
\centering
\includegraphics[height = 0.84\linewidth]{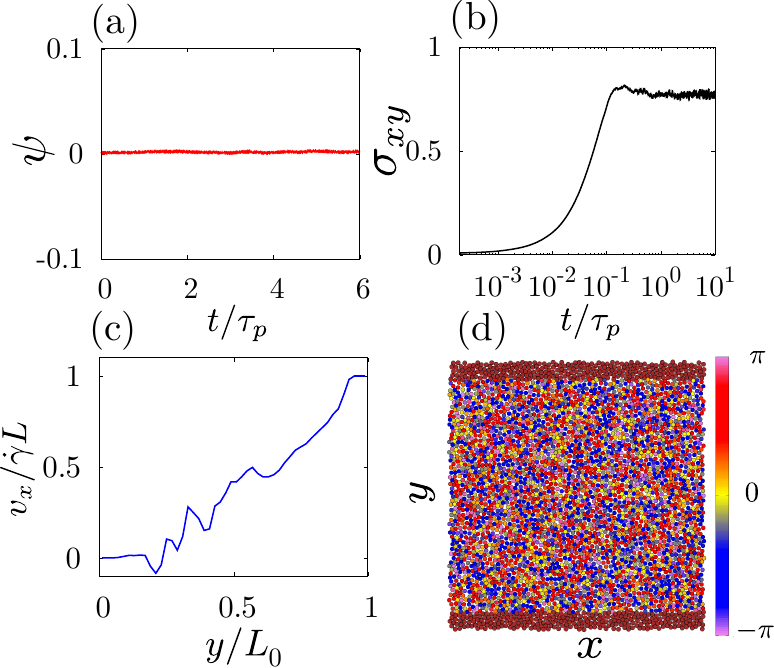}
\caption{Disordered phase (marked by red square in phase diagram in Fig.~\ref{fig:phased}(b)): (a) Orientational order parameter $\psi$ (see text for details) shows no emergent order in steady state. (b) Shear stress $\sigma_{xy}$ exhibits initial growth during startup of steady shear, with  fluctuations around a finite value in steady state. (c) Instantaneous velocity profile (see Subsection: Material and Methods for details) in steady state is essentially linear with small non-affine fluctuations. (d) Snapshot of the system in the steady state shows uniform distribution of particle orientation $\theta$ as shown by colour bar.}
\label{fig:disordered}
\end{figure}
Following the approach in Ref.~\cite{mandal20,mandalprl20} we keep number density high ($\rho=1.2$) and focus mainly on the athermal limit ($T=0$). We will see later, however, that for small persistence times, $\tau_p=O(1)$, the active system can be mapped to an equivalent passive thermal system. Note also that while our the equation of motion does contain an inertial term, the value of friction coefficient ($\gamma=1$) that we use ensures that the dynamics is mostly overdamped~\cite{rottler13}. We focus mostly on the two extreme limits of very small and very large persistence time $\tau_p$, choosing specifically $\tau_p=1$ and $\tau_p=10^3$. The key parameters for the phase diagrams we construct in this study are then the magnitude $f$ of the self-propulsion force and the imposed shear rate $\dot{\gamma}$.

\section*{Results}

\subsection*{Small Persistence Time} 

We first performed steady shear simulations in the small persistence time limit ($\tau_p=1$) of our model active glass. In this regime the behaviour can be understood using the idea of an effective temperature generated by the rapidly fluctuating active forcing. We measured the steady state shear stress $\langle \sigma_{xy} \rangle$ for different applied shear rates $\dot{\gamma}$ and compared the flow curve of the active glass with $\tau_p=1$ and $f=0.7$ to the flow curve of the corresponding passive system ($f=0$) at an equivalent effective temperature $T_{\rm eff}=0.1$. These two sets of flow curves match very well (as shown in supplementary figure Fig.~S1(b); see also SI for further discussion of the effective temperature). Indeed, from the time correlations of the active forces one expects $T_{\rm eff}$ to scale as $\tau_p f^2$ and this effective temperature description generally works well for active glasses in the small persistence time limit~\cite{mandal20,mandalprl20}. For most of the stress measurements in this study we use the virial stress, defined as $\sigma_{\alpha \beta}=-V^{-1}\sum^{N}_{i=1} r^\alpha_i F^\beta_i$ where $\mathbf{r}_i$ is the position of the $i$-th bulk particle, $\mathbf{F}_i$ is the total interaction force on this particle and $V$ is the volume of the bulk and $N$ is the number of particles in the bulk. We have also looked at the active stress defined as $\sigma^{a}_{\alpha \beta}=-V^{-1}\sum^{N}_{i=1} r^\alpha_i f^\beta_i$ where $\mathbf{f}_i=f\mathbf{n}_i$ is the active force acting on particle $i$. We observe that it fluctuates around zero (see supplementary Fig.~S1(c)) and so can be ignored in the average stress; its fluctuation amplitude depends strongly on the active force magnitude ($f$) but only very weakly on the shear rate ($\dot{\gamma}$), see supplementary Fig.~S1(c). As we are dealing with an active system, where the definition and meaning of stress has been much discussed, we also checked whether the virial stress calculated from the bulk matches with the direct measurement of stress from the forces on the walls. We see good agreement between the two approaches, both in a plot of stress versus strain as shear is started up at fixed shear rate, and in the flow curve ($\langle \sigma_{xy}\rangle$ versus $\dot{\gamma}$). We refer to the supplementary material for further discussion of the different stress definitions and their comparison (see in particular supplementary Fig.~S2). 

\subsection*{Large Persistence Time}

We observe the physically most interesting behaviour for large values of the persistence time, in particular with regards to spatial ordering of the orientation of the active particles (as given by the direction of their propulsive force). We therefore now fix the persistence time to the large value $\tau_p=10^3$ and vary active forcing $f$ and shear rate $\dot{\gamma}$ within a broad range of values $f=\{0,2.6\}$, $\dot{\gamma}=\{10^{-4},10^{-1}\}$. We find three type of dynamical phases (see Fig.~\ref{fig:phased}(b)): (a) disordered, (b) propulsion-induced ordered (PIO) and (c) shear-induced ordered (SIO). To differentiate these phases we use firstly an orientational order parameter. This is determined from the joint distribution $P(\theta,y)$ of the $y$-coordinates and orientations $\theta$ of the bulk particles as 
\begin{equation}
      \psi=\int_{-L/2}^{L/2}P(y) \phi(y) \sin{\left( \frac{2 \pi y}{L}\right)} dy  .
      \label{order1}
\end{equation}  
Here $\phi(y)$ is in turn defined as
\begin{equation}
 \phi(y)=\frac{\int^{\pi}_{-\pi}  P(\theta,y)\sin{\theta}\, d\theta}{P(y)},
 \quad P(y)=\int^{\pi}_{-\pi}  P(\theta,y) d\theta.
  \label{order2}
\end{equation}
Intuitively, $\phi(y)$ measures the dominant non-uniformity in the orientation distribution as a function of $y$ and is positive if particles point primarily towards the upper wall ($\theta\approx\pi/2$), and negative in the opposite case. The global order parameter $\psi$ thus detects whether the preferred particle orientation varies significantly in space, such that particles in the upper half of the system tend to point to the upper wall while those in the lower half point in the opposite direction. Both $\phi(y)$ and $\psi$ vanish for a uniform orientational distribution, $P(\theta,y)=P(y)/(2\pi)$. The colour bar in the phase diagram (see Fig.~\ref{fig:phased}(b)) in the $(\dot\gamma,f)$-plane shows the average value of the orientational order parameter $\psi$ in the steady state; the blue line is the contour for a small constant $\psi$ (we choose $\psi=0.05$) and so marks the boundary between orientationally ordered and disordered phases.

The black solid line and the white dashed line together mark the boundary between PIO and SIO, according to two complimentary criteria. The black solid line was calculated from the average stress and separates the regions of small and large stresses by a contour of constant $\langle \sigma_{xy} \rangle=0.4$ (roughly halfway between the typical values $\ll 1$ for PIO and $\approx 1$ for SIO). The white dashed line was determined from the strength of velocity fluctuations away from the affine flow field; it thus marks the boundary between phases with dominantly affine flow and those with significantly non-affine flow.
In particular, we measure the average velocity in the shear direction $\langle v_x(y)\rangle$ as a function of $y$ and define $v^{\rm NA}$ as the root-mean-squared deviation of this (see Materials and Methods for definition) from the affine flow velocity $v_{Fx}(y)$. The white dashed line in Fig.~\ref{fig:phased}(b) is then the contour ${v^{\rm NA}}/{v_w}=0.2$, where we use the upper wall velocity $v_w=\dot{\gamma}L$ as a natural velocity scale for the affine flow.

In the disordered phase the system behaves like a typical supercooled or glassy solid with Herschel-Bulkley rheology and a nonzero yield stress, around which stress fluctuations are seen in the steady state (see Fig.~\ref{fig:disordered}(b)). The particle orientations do not show any ordering (see Fig.~\ref{fig:disordered}(a) and the steady state snapshot in Fig.~\ref{fig:disordered}(d)) as $P(\theta,y)/P(y)$ remains close to ${1}/{2 \pi}$. The velocity profile remains linear with almost negligible non-affine fluctuations (see Fig.~\ref{fig:disordered}(c)). These small non-affine fluctuations increase with the active forcing $f$ but decrease with increasing shear rate $\dot{\gamma}$ (data not shown). These trends can be seen as the precursors of respectively the PIO phase, which appears at larger $f$ and exhibits significantly non-affine velocities as we discuss next, and the SIO phase at larger $\dot\gamma$ with its essentially affine flow.

\begin{figure}
\centering
\includegraphics[width = \columnwidth]{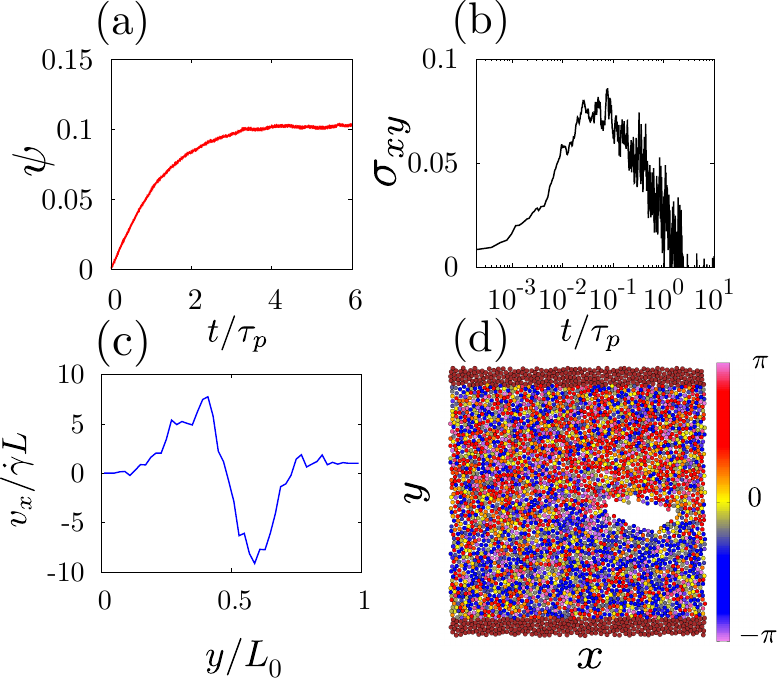}
\caption{Propulsion-induced ordered  phase (cyan circle in phase diagram in Fig.~\ref{fig:phased}(b)): (a) Orientational order parameter $\psi$ shows growth during startup of steady shear, with saturation for  $t\approx \tau_p$. (b) Shear stress $\sigma_{xy}$ fluctuates around zero in the steady state. (c) Instantaneous velocity profile in steady state demonstrates very strong non-affine fluctuations. (d) Snapshot of the system shows weak orientational order (see colour bar) and strong density inhomogeneities including transient cavity formation.}
\label{fig:pio}
\end{figure}

Upon increasing $f$ at small shear rate $\dot{\gamma}$ one enters the propulsion-induced ordered phase. This is characterized by a moderate amount of orientational ordering (see Fig.~\ref{fig:pio}(a,d)) but strong non-affine flows created by the ``stirring'' arising from the strong active forces (see Fig.~\ref{fig:pio}(c)). The system also shows significant density inhomogeneities ranging up to the formation of  transient cavities (see supplementary movie) that remodel dynamically. The spatial segregation into dense regions and cavities with almost zero density can be thought of as an inverted type of motility-induced phase separation (MIPS)~\cite{nardini18,nardini20,leticia20}. Because of this dynamic cavity formation and strong internal flows during the steady shear, the system is unable to sustain any significant shear stress in the steady state: the average stress reaches very low values, with strong fluctuations around the mean (see Fig.~\ref{fig:pio}(b)). We call this phase ``propulsion-induced ordered'' because the spatial segregation is observable even for $\dot{\gamma} \to 0$, where the motion of the particles and hence the overall physics is dominated by the active propulsion forces.

\begin{figure}
\centering
\includegraphics[width=\columnwidth]{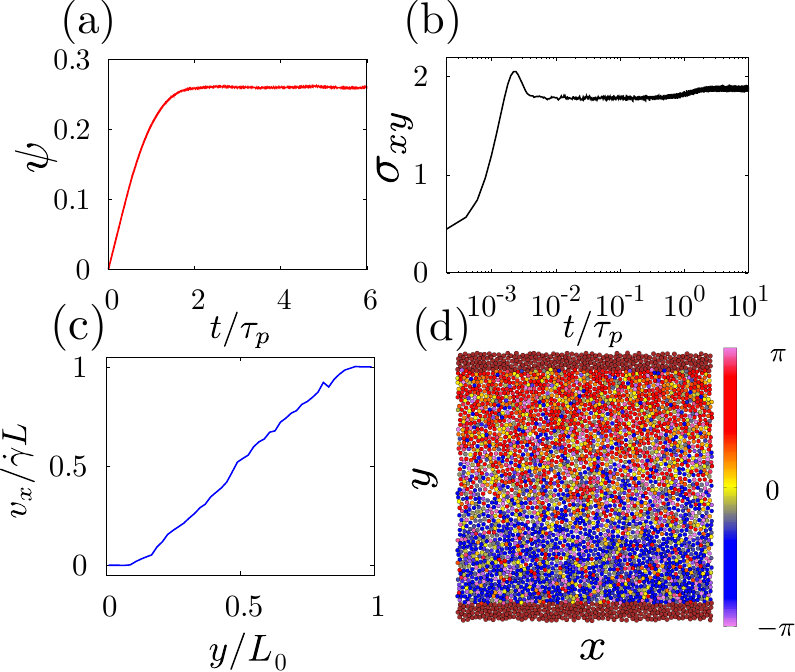}
\caption{Shear Induced Ordered  phase (marked by green triangle in phase diagram in Fig.~\ref{fig:phased}(b)): (a) Orientational order parameter $\psi$ indicates a strongly ordered steady state. (b) Shear stress remains finite in steady shear, reaching a constant value at $t\approx \tau_p$. (c) Instantaneous velocity profile at steady state shows linear velocity profile with negligible non-affine fluctuations. (d) Snapshot of the system shows strong orientational order (see colour bar) but essentially  homogeneous density.}
\label{fig:sio}
\end{figure}

We next consider the shear-induced ordered phase. This can also be entered from the disordered phase, but along a different route: starting at moderate active forcing, $f \sim 1$, in the disordered phase one increases the shear rate $\dot{\gamma}$ from a very low value $\sim 10^{-3}$ to a quite substantial rate $\sim 10^{-1}$. The resulting SIO phase shows liquid-like properties in many respects. For example, it has an almost linear velocity profile with negligible non-affine fluctuations (see Fig.~\ref{fig:sio}(c)). The shear stress increases after shear startup and reaches a constant steady state value on time scales around or above $\tau_p$ (see Fig.~\ref{fig:sio}(b)). Surprisingly, however, these liquid-like features are combined with strong spatial order in the particle orientations (see Fig.~\ref{fig:sio}(a,d)). 

Summarizing our findings for the properties of the two ordered phases, PIO and SIO, these differ substantially in the spatial uniformity of the number density, in the presence or absence of non-affine flow and in their capability to support significant shear stress. The PIO phase is dominated by the physics of phase segregation, and shear does not play a significant role in its formation or stability. The SIO phase, on the other hand, exists only at reasonably large shear rates and active propulsion does not on its own guarantee the existence or stability of such a phase.

We next show that the qualitative behaviour of the SIO phase as an  affinely flowing yet orientally ordered state can be understood by a simple analytical approach. This is based on an effective Fokker-Planck description of the single particle dynamics and will allows us to make quantitative  predictions that can be verified by our simulations.

\subsection*{Effective single particle description}

The dynamics of our model glass can in principle be described by a Fokker-Planck equation for the $3N$ particle coordinates and orientations $(x_i,y_i,\theta_i)$,  together with the associated momenta and  supplemented by terms describing the wall motion. As the qualitative physics we are describing corresponds to an overdamped limit, the momenta can be ignored to a good approximation. The equation of motion for a single particle can then formally be derived by integrating out all other degrees of freedom~\cite{zwanzig01}. We neglect the memory effects that generically occur in such a reduction (see e.g.~\cite{rubin14,herrera20}) and write an approximate  Fokker-Planck equation for the joint distribution $P(\theta,y)$ of the orientation $\theta$ and $y$-position of a single particle:
\begin{equation}
\frac{\partial P(\theta,y)}{\partial t} +\mu f \sin{\theta} \frac{\partial P(\theta,y)}{\partial y}= D \frac{\partial^2 P(\theta,y)}{\partial y^2} + D_{\theta} \frac{\partial^2 P(\theta,y)}{\partial \theta^2}
\label{fpe}
\end{equation}
Here $\mu$ is the effective mobility, $f$ is the active force magnitude as before, $D$ is the effective diffusion constant and $D_\theta=1/\tau_p$ is the orientational diffusion constant. The second term on the left hand side represents advection of probability due to active forcing, with the average $y$-velocity of the particle being proportional to the $y$-component $f\sin \theta$ of the active force. The terms on the right hand side describe translational and rotational diffusion, respectively. Our aim is now to obtain the steady state solution $P(\theta,y)$, from which predictions for the orientational order parameter can be derived according to \eqref{order1} and \eqref{order2}. A Fourier decomposition of the angular dependence gives \begin{equation}
P(\theta,y)=\sum_{m=-\infty}^{\infty} e^{i m \theta} P_{m}(y)
\label{mode}
\end{equation}
where $P_m(y)$ is the $y$-dependent coefficient of the $m$-th Fourier mode. Substituting the representation \eqref{mode} into \eqref{fpe} we obtain a system of coupled equations for the $P_m(y)$:
\begin{equation}
\frac{\mu f}{2 i} \left[ \frac{\partial P_{m-1}(y)}{\partial y}-   \frac{\partial P_{m+1}(y)}{\partial y} \right]     =
 D  \frac{\partial^2 P_{m}(y)}{\partial y^2} -  m^2 D_{\theta}   P_{m}(y).
\end{equation}
To find a closed form approximation for $P(\theta,y)$ we use a Galerkin truncation~\cite{fletcher84}; specifically we truncate the series in \eqref{mode} after the leading terms with $m=0,\pm 1$. We have checked that the inclusion of higher modes does not change the predictions qualitatively (see supplementary Fig.~S3(a,b)). The boundary conditions required to fix the solution for $P(\theta,y)$ arise from the fact that the upper and lower walls in our system are impermeable so that the probability current in the $y$-direction, which is given by
\begin{equation}
J_y=(\mu f \sin{\theta})P-D\frac{\partial P}{\partial y}
\end{equation}
has to vanish there: 
\begin{figure}
\centering
\includegraphics[width=1.02\columnwidth]{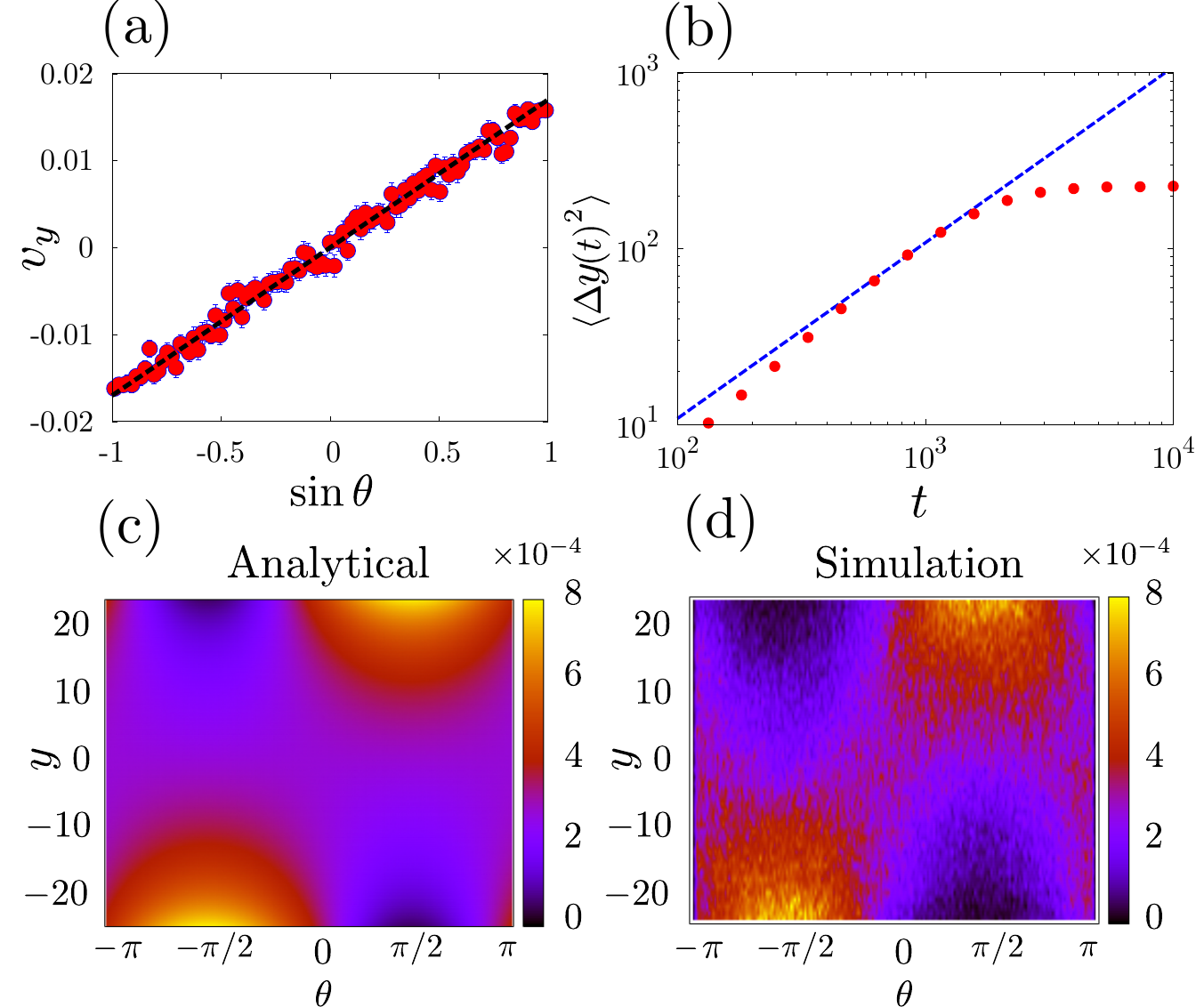}
\caption{Single particle parameters in SIO phase marked in Fig.~\ref{fig:phased}(b) (open triangle) (a) Average velocity of particles around $y=0$ in shear gradient ($y$) direction, plotted against $\sin\theta$, giving a linear relation as assumed in our single particle theory; the slope identifies the mobility $\mu$. (b) Mean-squared displacement of particles starting around $y=0$ (marked by red points) and a linear fit (blue dashed line) to extract $D_m$.  (c,d) Comparison of steady state $P(\theta,y)$ from the analytical single particle theory and from simulation, showing good agreement without fit parameters.}
\label{fig:compare}
\end{figure}
\begin{equation}
  J_y = 
  \begin{cases}
    0, & \text{at } y = -L/2 \\
    0, & \text{at } y = +L/2 
  \end{cases}
\end{equation}
Away from the walls we find that $J_y$ is generically nonzero, reflecting the non-equilibrium character of the system. The full current vector $(J_\theta,J_y)$ exhibits interesting structure, with the corresponding flow splitting the $(\theta,y)$-plane into a number of distinct regions as shown in supplementary Fig.~S3(c). With the above boundary conditions we find the steady state solution for the dominant Fourier modes as
\begin{equation}
 P_0(y)=\frac{1}{2 \pi}\frac{ \cosh(\alpha y)+\zeta \cosh(\alpha L/2)}{\zeta L \cosh(\alpha L/2) + \frac{2}{\alpha} \sinh(\alpha L/2)}\ ,   
\end{equation}
and
\begin{equation}
P_1(y)=-i \frac{\alpha D}{\mu f} \frac{1}{2 \pi} \frac{\sinh(\alpha y)}
{\zeta L \cosh(\alpha L/2) + \frac{2}{\alpha} \sinh(\alpha L/2)}
%P_0(y) \phi(y)
\label{P1}
\end{equation}
with $P_{-1}(y)=-P_{1}(y)$. 
Here the inverse length scale $\alpha$ is defined via $\alpha^2=(\mu^2 f^2+ 2 D D_\theta)/(2 D^2)$. Inserting into the Fourier expansion of~\eqref{mode} with our Galerkin truncation 
\begin{equation}
P(\theta,y)=e^{i \theta} P_1(y) + e^{-i \theta} P_{-1}(y) + P_0(y) 
\end{equation}
gives the desired steady state joint distribution 
\begin{equation}
\begin{aligned}
\begin{split}
P(\theta,y)=& 
\frac{\alpha D}{\mu f} \ \frac{2 \sin{\theta} \sinh(\alpha y)}{2 \pi (\zeta L \cosh(\alpha L/2) + \frac{2}{\alpha} \sinh(\alpha L/2))} \\
& {}+{}\frac{ \cosh(\alpha y)+\zeta \cosh(\alpha L/2)}{2 \pi (\zeta L \cosh(\alpha L/2) + \frac{2}{\alpha} \sinh(\alpha L/2))}
\end{split}
\end{aligned}
\end{equation}
From this we can finally deduce 
the local orientational order parameter $\phi(y)$ defined in~\eqref{order2}:
\begin{equation}
    \phi(y)=\frac{\alpha D}{\mu f} \frac{\sinh(\alpha y)}{\cosh(\alpha y)+ \zeta \cosh(\alpha L/2)}
\end{equation}

To assess these theoretical predictions we can take most parameters directly from simulations and use for a  typical SIO phase $f=1$, $\tau_p=10^3$ and $L_0=50$. The remaining two effective single particle parameters can be measured directly from the simulation (see Fig.~\ref{fig:compare}(a,b)): the mobility $\mu=v_y/(f \sin{\theta})$  can be extracted from the relation between the average velocity of the particles and their orientation and the diffusion constant from $D_m=\langle {\Delta y(t)}^2 \rangle/(2t)$ (see Material and Methods). As the diffusive description is approximate (see below) we take a somewhat larger $D$ in the theory, $D=3.3 D_m$. With $\mu$ and $D$ fixed independently, there are no remaining free parameters in the theory. Fig.~\ref{fig:compare}(c,d) shows the comparison of the theoretically predicted $P(\theta,y)$ and the steady state simulation data in the corresponding SIO phase. We observe very good qualitative and semi-quantitative agreement, particularly given the approximations inherent in the theory: (i) the simulations show a super-diffusive growth of the mean-squared particle displacement, suggesting that memory effects cannot be fully neglected as we have done;  (ii) the effective diffusivity $D$ and mobility $\mu$ may exhibit some dependence on density and therefore $y$, whereas we have taken these parameters as constant. Overall, our simple single particle theory offers a remarkably good description of the SIO phase and its orientational ordering. Within the theoretical picture the degree of orientational ordering is governed by the competition between the orientational bias of the particle velocities in the shear gradient direction on the one hand, and the disordering tendency of the rotational diffusion on the other. Ordering is observable when the steady shear leads to a sufficiently large effective mobility and hence a stronger tendency towards orientational order.

\subsection*{Discussion}

In this work we explored the effect of steady shear deformation in a model active glassy material and report a new type of orientational ordering that is facilitated by shear. Using different physical quantities including non-affinity in the velocity profile, steady state shear stress and an orientational order parameter, we were able to distinguish a number of qualitatively distinct dynamical steady states: (i) a disordered phase,  (ii) a propulsion-induced ordered phase and (iii) a shear induced-ordered phase. We observe in particular that shear can help the system to order  orientationally, effectively by mobilizing particles sufficiently to follow their orientational bias. For the SIO phase, which exhibits strong orientational ordering yet essentially affine flow, we  constructed a single particle Fokker-Planck theory that predicts the joint steady state distribution $P(\theta,y)$ and hence the degree of orientational order in the system. The predictions compare well with simulation data. There are no free fit parameters in this comparison as the two required effective single particle quantities (mobility and diffusion constant) can be measured directly in the simulations.

Shear-induced phase transitions and ordering phenomena in {\em passive} systems have been an active topic of research for some time~\cite{ackerson88,ackerson90,stevens93,wu09}. It has been shown that although shear can destroy order by melting a system~\cite{stevens93}, it can also induce ordering in the sense of crystallisation~\cite{ackerson88,stevens93,wu09,fardin17,zaccarelli18} or alignment for elongated particles~\cite{koppi93,szabo12}. Shear-induced crystallization has also been reported very recently in passive colloidal Janus particles~\cite{zihan19}. Our results demonstrate that new forms of shear-induced ordering can arise in active systems, which are inherently out of equilibrium even without shear. This will add new directions to the exciting paradigm of ordering through steady shear driving. 

It will be an interesting challenge to see whether our theoretical approach can be extended into a complete theory for all phases we see, including a prediction of the phase boundaries. This would require in particular accounting appropriately with the spatial inhomogeneities of the PIO phase. Another interesting direction would be to study shear ordering in chiral active systems, where qualitatively new phenomena might be expected. The novel orientational ordering we have found can also be explored and exploited in synthetic active glassy systems, in controlled experiments on active colloids or active granular matter.

\subsection*{Materials and Methods}
\label{mandm}

The simulations were performed in two spatial dimensions with a square box with periodic boundary condition implemented along the direction of flow. We used modified Langevin dynamics~\cite{beard00} for the MD simulation with $dt=0.002$. We averaged all steady state quantities over a time scale of $t=10^4$ after first allowing the same amount of time for the system to reach a steady state. This is ample as for our largest $\tau_p=10^3$ the relaxation to the steady state still happens on a timescale of order $\tau_p$. Apart from this temporal averaging, the order parameter $\psi$, shear stress $\sigma_{xy}$, and other relevant quantities are also averaged over $128$ independent simulations. We compared simulations with $N_0=1000$ and $N_0=4000$ particles to check for finite size effects; all results shown were generated for $N_0=4000$. Here $N_0$ includes both the bulk and boundary (wall) particles. This corresponds to a box size of $L_0=57.73$ (as number density is $\rho=1.2$) with a distance between the two walls of $L=50$ and an wall thickness of approximately $4 \sigma_{AA}$.  Velocity profiles are calculated by dividing the whole system into $N_s=50$ slabs in the $y$-direction. For the calculation of $v^{NA}$ we average the mean square fluctuations around the affine velocity $v_{Fx}(y)$ across $y$-bins and evaluate the root of this mean squared fluctuation. This means that for $N_0=4000$, averages within a slab are taken over $80$ particles on average. To determine the measured effective diffusion constant $D_m$ we divided the system into $5$ slabs and measured $\langle (\Delta {y(t)})^2 \rangle$ in each of them. The diffusion constant was then determined from $\langle(\Delta y(t))^2\rangle/(2t)$ at the time $t$ where $\langle (\Delta y(t))^2\rangle^{1/2} \sim 10$,  which is the size of each slab in the $y$-direction.

We are grateful to J\"org Rottler and Rohit Jain for insightful discussions. This project has received funding from the European Union’s Horizon 2020 research and innovation programme under Marie Sk\l odowska-Curie grant agreement No.\ 893128.

%\showacknow{}

\bibliography{bibfile_rheact}

\end{document}

\title{Supplementary Information: Shear Induced Orientational Ordering in Active Glass}

\author{Rituparno Mandal}
\affiliation{Institute for Theoretical Physics, Georg-August-Universit\"{a}t G\"{o}ttingen, 37077 G\"{o}ttingen, Germany.}
\author{Peter Sollich}
\affiliation{Institute for Theoretical Physics, Georg-August-Universit\"{a}t G\"{o}ttingen, 37077 G\"{o}ttingen, Germany.}
\affiliation{Department of Mathematics, King's College London, London WC2R 2LS, UK}

\maketitle

\renewcommand{\thefigure}{S\arabic{figure}}

\subsection*{Small persistence time ($\tau_p \to 0$)}

\begin{figure}
\centering
\includegraphics[width=1.\columnwidth]{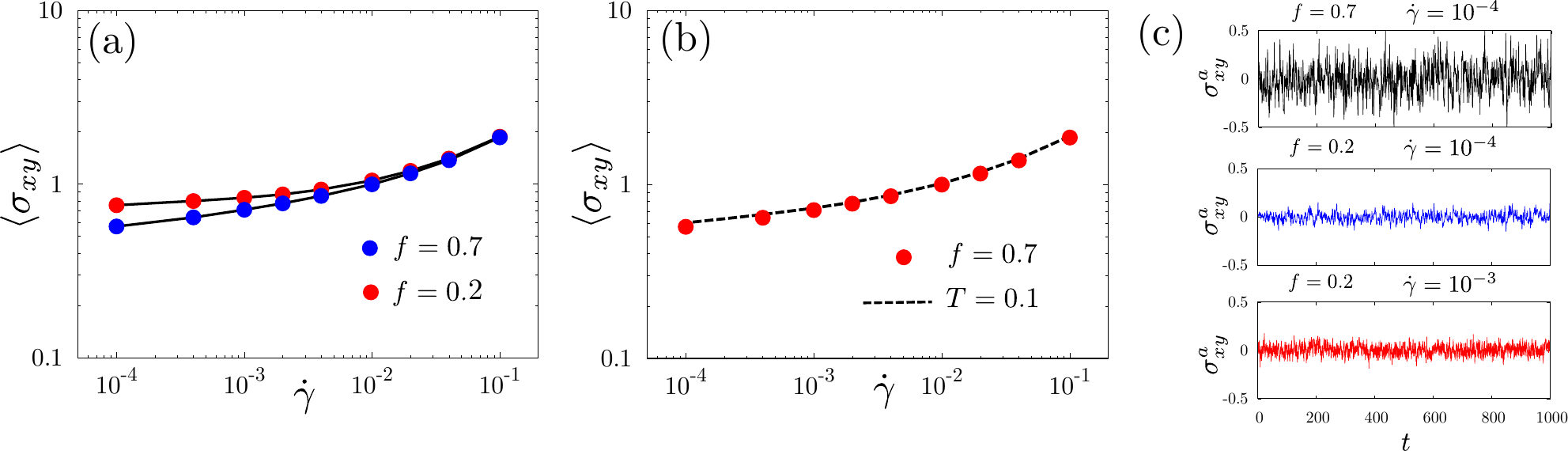}
\caption{(a) Flow curve (passive virial shear stress versus strain rate) for two different strengths $f$ of active forcing shows Hershel-Buckley yield-stress behaviour, with the yield stress decreasing with active forcing. (b) A comparison of the flow curve for $f=0.7$ and $\tau_p=1$ with that of a thermal system with the corresponding effective temperature $T_{\rm eff} = 0.1 $.
(c) The time series of active stress (see MS for definition) fluctuates around zero where the fluctuations are weakly dependent on the applied shear rate (see lower two panels) but increases strongly with the increasing active forcing $f$ (see upper two panels).}
\label{fig:smalltaup}
\end{figure}

Shear simulations of the active glass model were run at a fixed small value of the persistence time, $\tau_p=1$, for different values of the active forcing $f$. For small $f$, {\textit{e.g.}}\ $f=0.2$, the flow curve (plot of average steady shear stress $\langle \sigma_{xy} \rangle$ versus shear rate $\dot{\gamma}$) reveals that the system behaves like a yield-stress fluid of Herschel–Bulkley type. Here the angular brackets $\langle \ldots\rangle$ represent a steady state average over time and also over $128$ independent simulation runs. Increasing the active forcing $f$ decreases the yield stress (see Fig.~\ref{fig:smalltaup}(a)), an effect that is analogous to that of a temperature increase for sheared passive systems~\cite{joyjit10,ikeda12,wyart20}.

Now, in the limit $\tau_p \to 0$, the behaviour of the system can be mapped to that of a passive system with effective temperature $T_{\rm eff}=c f^2 \tau_p$ where $c \sim 0.2$~\cite{mandal20,mandalprl20}. To verify this in the rheological scenario we performed two different sets of simulations:  
(a) the active glass for $\tau_p=1$ and $f=0.7$, (b) a passive system whose temperature is same as the effective temperature $T_{\rm eff}= 0.1$ of the active system. The comparison between the flow curves (see Fig.~\ref{fig:smalltaup}(b)) clearly establishes that the rheological response of the active system can be understood by considering a passive system at the equivalent effective temperature.

All the analysis described above was performed using the bulk virial stress, which that does explicitly contain any contribution from the active forces. Indirectly, the active forces of course affect what configurations the active systems visits during its non-equilibrium dynamics and therefore what its bulk virial stress is. We have verified explicitly that the active force term in the stress term averages to zero and so does not contribute to the flow curve. In Fig.~\ref{fig:smalltaup}(c) we show this explicitly in terms of the time series of the $xy$-component of the active stress, $\sigma^a_{xy}$, which fluctuates around zero. The amplitude of these fluctuations depends weakly on shear rate but increases rapidly with increasing active forcing $f$ (Fig.~\ref{fig:smalltaup}(c)) as the PIO phase is approached.

\subsection*{Stress comparison}

It was a matter of debate for some time whether the stress tensor is a well-defined quantity in an out-of-equilibrium system made up of active particles. Recent studies~\cite{solon15,solonprl15} have shown that, if there is no torque acting between the particles (arising from some kind of aligning interaction) or between the particles and the wall, then the pressure (or stress) is a well-defined quantity. The kind of model we are considering imposes no torque on the active particles from either particle-particle or particle-wall interactions. Therefore we expect that stress will be a well-defined thermodynamic variable and different definitions of stress will agree with each other. 

\begin{figure}
\centering
\includegraphics[width=1.\columnwidth]{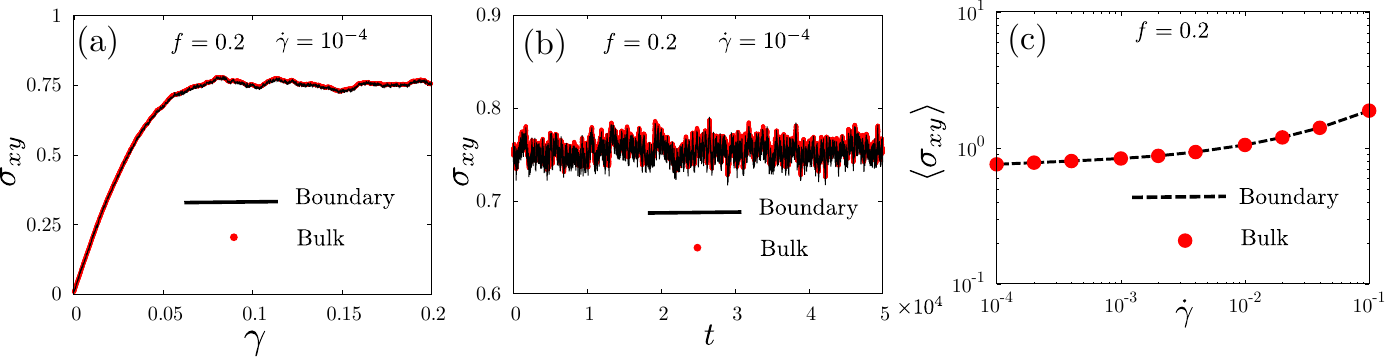}
\caption{(a) Stress $\sigma_{xy}$ versus strain $\gamma$ in shear start up: (i) virial stress calculated from the bulk (red points) and (ii) stress determined via the force on the wall (black solid line); (b) same as (a) but plotted against $t$ and in steady state; (c) the flow curve ($\langle \sigma_{xy} \rangle$ vs $\dot{\gamma}$) calculated using both definitions of the stress (red points: bulk virial stress, black dashed line: wall force stress), showing perfect agreement.
}
\label{fig:wallbulk}
\end{figure}

To verify this conclusion explicitly, we calculated the stress using two different definitions, (a) bulk stress, calculated using the virial formula and (b) wall stress, calculated from the net force exerted on the wall. In Fig.~\ref{fig:wallbulk}(a) and (b) we show that the stress versus strain ($\sigma_{xy}$ vs $\gamma$) plots match exactly when we compare these two stress definitions. The simulations were performed for $\dot{\gamma}=10^{-4}$ at $f=0.2$ and $\tau_p=1.0$. We further compared (see Fig.~\ref{fig:wallbulk} (c)) the flow curve ($\langle \sigma_{xy} \rangle$ vs $\dot{\gamma}$) for the two definitions and they also match perfectly. As mentioned before, the angular brackets $\langle \ldots\rangle$ represent a steady state average over time and also over $128$ independent simulation runs.

\subsection*{Fokker-Planck Equation and Solution}

\begin{figure}
\centering
\includegraphics[width=1.\columnwidth]{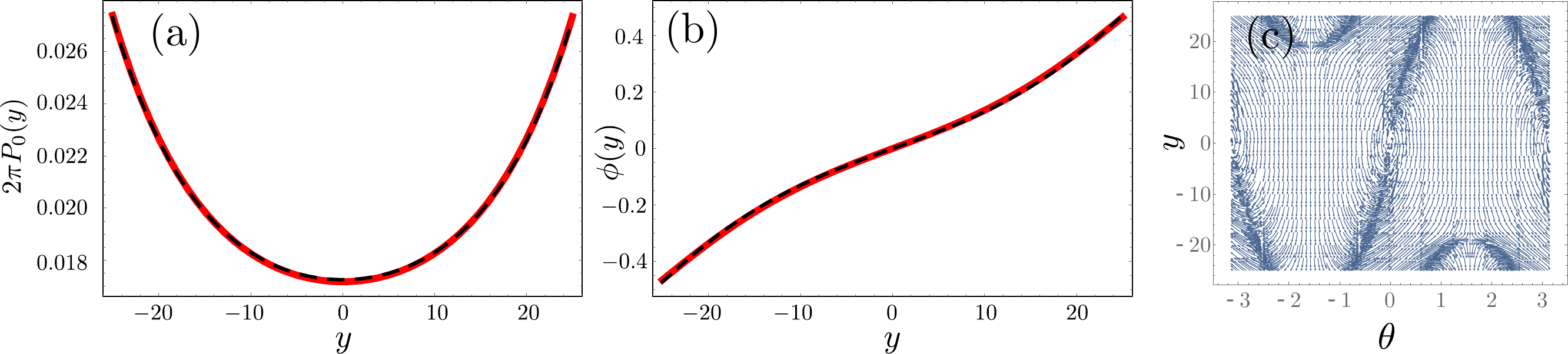}
\caption{(a) The $0$-th Fourier mode plotted as a function of $y$ where the red line represents the solution using $m=0,\pm 1$ modes and the black dashed line shows the solution using $m=0,\pm 1, \pm 2$. (b) The $y$ dependent orientational order parameter $\phi(y)$ plotted as a function of $y$ for $m=0,\pm 1$ (red line) and $m=0,\pm 1, \pm 2$ (black dashed line). Both (a) and (b) demonstrate that a three mode calculation provides sufficient accuracy. (c) Flow lines of the full steady state current vector $(J_\theta,J_y)$ plotted in the $\theta$, $y$ plane, as derived from  the solution given in~\eqref{final}.}
\label{fig:higherorder}
\end{figure}

The Fokker-Planck equation for the joint probability distribution of the orientation of the active forcing ($\theta$) and the particle position ($y$) along the shear gradient direction can be written as
\begin{equation}
\frac{\partial P(\theta,y)}{\partial t} +\mu f \sin{\theta} \frac{\partial P(\theta,y)}{\partial y}= D \frac{\partial^2 P(\theta,y)}{\partial y^2} + D_{\theta} \frac{\partial^2 P(\theta,y)}{\partial \theta^2}
\end{equation}
where $\mu$ is the effective mobility, $f$ is the magnitude of active forcing, $D$ is the effective diffusion constant and $D_\theta=1/\tau_p$ is the orientational diffusion constant of the active forcing. In steady state this gives
\begin{equation}
\mu f \sin{\theta} \frac{\partial P(\theta,y)}{\partial y}= D \frac{\partial^2 P(\theta,y)}{\partial y^2} + D_{\theta} \frac{\partial^2 P(\theta,y)}{\partial \theta^2}.
\label{steady}
\end{equation}
By Fourier transform in $\theta$ we can write the solution in the following form:
\begin{equation}
P(\theta,y)=\sum_{m=-\infty}^{\infty} e^{i m \theta} P_{m}(y)
\end{equation}
Using this Fourier expansion we obtain from~\eqref{steady}
\begin{equation}
 \sum_{m=-\infty}^{\infty} \left[ \frac{\mu f}{2 i} e^{i m \theta} \frac{\partial P_{m-1}(y)}{\partial y}- \frac{\mu f}{2 i} e^{i m \theta} \frac{\partial P_{m+1}(y)}{\partial y} \right]     =
 \sum_{m=-\infty}^{\infty} D e^{i m \theta} \frac{\partial^2 P_{m}(y)}{\partial y^2} - \sum_{m=-\infty}^{\infty} m^2 D_{\theta} e^{i m \theta}  P_{m}(y).
\end{equation}
This further leads to (for each $m$)
\begin{equation}
  \frac{\mu f}{2 i} \left[ \frac{\partial P_{m-1}(y)}{\partial y}-   \frac{\partial P_{m+1}(y)}{\partial y} \right]     =
 D  \frac{\partial^2 P_{m}(y)}{\partial y^2} -  m^2 D_{\theta}   P_{m}(y).
\end{equation}
Let us now truncate the series by assuming $P_m(y)=0$ for $|m|\ge 2$. Fig.\ref{fig:higherorder} (a,b) shows that inclusion of higher order modes does not change the results qualitatively in the parameter range relevant for the SIO phase. For $m=0$ we have
\begin{equation}
  \frac{\mu f}{2 i} \left[ \frac{\partial P_{-1}(y)}{\partial y}-   \frac{\partial P_{1}(y)}{\partial y} \right]     =
 D  \frac{\partial^2 P_{0}(y)}{\partial y^2}  
 \label{modep0}
\end{equation}
while for $m=\pm 1$
\begin{equation}
  \frac{\mu f}{2 i} \left(\pm  \frac{\partial P_{0}(y)}{\partial y} \right)    =
 D  \frac{\partial^2 P_{\pm{1}}(y)}{\partial y^2} -  D_{\theta}   P_{\pm 1}(y).
 \label{modepm}
\end{equation}
From~\eqref{modepm} we obtain
\begin{equation}
P_1(y)=-P_{-1}(y)
\label{modepm1}
\end{equation}
and combining~\eqref{modep0} and~\eqref{modepm1}\ we get
\begin{equation}
    \frac{\partial P_{1}(y)}{\partial y} =
-\frac{i D}{\mu f}  \frac{\partial^2 P_{0}(y)}{\partial y^2} 
\label{mode1}
\end{equation}
A derivative w.r.t.\ $y$ (of~\eqref{modepm}) then yields
\begin{equation}
  \frac{\mu f}{2 i}  \frac{\partial^2 P_{0}(y)}{\partial y^2}      =
 D  \frac{\partial^2 }{\partial y^2} \left( \frac{\partial P_{1}(y)}{\partial y}\right)-  D_{\theta}   \frac{\partial P_{1}(y)}{\partial y}.
\end{equation}
Now using~\eqref{mode1} we obtain
\begin{eqnarray}
  \frac{\mu f}{2 i}  \frac{\partial^2 P_{0}(y)}{\partial y^2}      =
 -\frac{iD^2}{\mu f}  \frac{\partial^4 P_0(y)}{\partial y^4}+  \frac{iD_{\theta} D}{\mu f}  \frac{\partial^2 P_{0}(y)}{\partial y^2}\\
 \frac{\partial^4 P_0(y)}{\partial y^4}= \left(\frac{\mu^2 f^2}{2D^2}+ \frac{D_{\theta}}{D}  \right)  \frac{\partial^2 P_{0}(y)}{\partial y^2}  
\end{eqnarray}
Writing $p=\frac{\partial^2 P_0(y)}{\partial y^2}$ and $\alpha^2=\frac{\mu^2 f^2+ 2 D D_\theta}{2 D^2}$ we find
that this linear differential equation has solutions of the form $e^{ky}$ with
$k=\pm \alpha$, so that $p=a_0 e^{\alpha y} + a_1 e^{-\alpha y}$. Integrating twice w.r.t.\ $y$ then leads to
\begin{equation}
 P_0(y)=\frac{a_0}{\alpha^2} e^{\alpha y} + \frac{a_1}{\alpha^2} e^{-\alpha y} +a_2 y+ a_3
 \label{modep02}
\end{equation}
and using~\eqref{mode1} we also get
\begin{equation}
 P_1(y)=-\frac{iD}{\mu f \alpha} \left[a_0 e^{\alpha y} - a_1 e^{-\alpha y}\right]+a_4.
\end{equation}
Thereafter using~\eqref{modep0} one can obtain
\begin{equation}
\frac{\mu f}{2 i} \left[\frac{a_0}{\alpha} e^{\alpha y} - \frac{a_1}{\alpha} e^{-\alpha y} +a_2\right]=D \alpha^2 \left(-\frac{iD}{\mu f \alpha}\right) \left[a_0 e^{\alpha y} - a_1 e^{-\alpha y} \right]+\frac{i D_{\theta}D}{\mu f \alpha}\left[a_0 e^{\alpha y} - a_1 e^{-\alpha y} \right]- D_{\theta} a_4
\end{equation}
which determines $a_4$ and so gives us
\begin{equation}
 P_1(y)=-\frac{iD}{\mu f \alpha} \left[a_0 e^{\alpha y} - a_1 e^{-\alpha y}\right]+\left(\frac{\mu f i}{2 D_{\theta}}\right)a_2
  \label{modep12}
\end{equation}
We now exploit the
boundary condition on $J_y$
\begin{equation}
  J_y = 
  \begin{cases}
    0, & \text{at } y = -L/2 \\
    0, & \text{at } y = +L/2 
  \end{cases}
\end{equation}
where $J_y$ is the $y$-component of the probability current with the form
\begin{equation}
J_y=(\mu f \sin{\theta})P-D\frac{\partial P}{\partial y}.
\label{currydef}
\end{equation}
Following the reasoning above we switch to the Fourier representation
\begin{equation}
J_y=\sum_{m=-\infty}^{\infty} \left[ \frac{\mu f}{2i}   P_{m-1}(y)- \frac{\mu f}{2i} P_{m+1}(y) -D   \frac{\partial P_m(y)}{\partial y}\right] e^{im \theta}
\end{equation}
The boundary condition therefore produces constraints allowing us to evaluate the constants of integration $a_0,a_1,a_2$ and $a_3$. We find $a_0=a_1$, $a_2=0$ and $a_3=2a_0 \cosh{\left(\frac{\alpha L}{2}\right)} \left[\frac{2 D^2}{\mu^2 f^2}-\frac{1}{\alpha^2}\right]$.
With~\eqref{modep02} and~\eqref{modep12} this yields
\begin{equation}
    P_0(y)=\frac{2a_0}{\alpha^2} \cosh{(\alpha y)} + 2a_0 \left[\frac{2D^2}{\mu^2 f^2}-\frac{1}{\alpha^2}\right] \cosh{(\alpha L/2)}
    \label{prefact}
\end{equation}
and 
\begin{equation}
    P_1(y)=-\frac{2 i D a_0}{\mu f \alpha} \sinh{(\alpha y)}
\end{equation}
We simplify the second prefactor in~\eqref{prefact} to
\begin{equation}
    \frac{2D^2}{\mu^2 f^2}-\frac{1}{\alpha^2}= \frac{2D^2}{\mu^2 f^2}-\frac{2 D^2}{\mu^2 f^2+ 2 D D_{\theta}}\\
    = \frac{2D^2}{\mu^2 f^2}\frac{2 D D_{\theta}}{2 \alpha^2 D^2}\\
    =\frac{\zeta}{\alpha^2}
\end{equation}
with $\zeta=\frac{2 DD_{\theta}}{\mu^2 f^2}$, so that $P_0(y)$ can be written more compactly as
\begin{equation}
    P_0(y)=\frac{2a_0}{\alpha^2} \left[ \cosh(\alpha y)+\zeta \cosh(\alpha L/2)\right]
\end{equation}
The normalisation $P(\theta,y)$ does not get any contribution from $P_{\pm 1}(y)$ as that part goes to zero because of the $\sin{\theta}$ factor. Hence the normalisation of $P_0(y)$ gives 
\begin{equation}
    \int_{-L/2}^{L/2} P_0(y) dy=\frac{1}{2 \pi}
\end{equation}
and this leads to
\begin{equation}
 P_0(y)=\frac{ \cosh(\alpha y)+\zeta \cosh(\alpha L/2)}{2 \pi (\zeta L \cosh(\alpha L/2) + \frac{2}{\alpha} \sinh(\alpha L/2))}    
\end{equation}
In our truncated form the joint distribution of $\theta$ and $y$ is then
\begin{equation}
P(\theta,y)=e^{i \theta} P_1(y) + e^{-i \theta} P_{-1}(y) + P_0(y) 
\end{equation}
We have finally
\begin{equation}
P(\theta,y)=2 \sin{\theta} \left(\frac{D \alpha }{\mu f}\right) \left(\frac{\sinh(\alpha y)}{2 \pi (\zeta L \cosh(\alpha L/2) + \frac{2}{\alpha} \sinh(\alpha L/2))} \right)+ \frac{ \cosh(\alpha y)+\zeta \cosh(\alpha L/2)}{2 \pi (\zeta L \cosh(\alpha L/2) + \frac{2}{\alpha} \sinh(\alpha L/2))}
\label{final}  
\end{equation}
where as mentioned before $\alpha=\sqrt{\frac{\mu^2 f^2 \tau_p+ 2 D}{2 D^2 \tau_p}}$ and $\zeta=\frac{2 D}{\mu^2 f^2 \tau_p}$. As in the main text we define the $y$-dependent orientational ordering
\begin{equation}
    \phi(y)=\frac{\int^{\pi}_{-\pi} \sin{\theta} P(\theta,y) d\theta}{\int^{\pi}_{-\pi}  P(\theta,y) d\theta}
\end{equation}
which becomes explicitly
\begin{equation}
    \phi(y)=\frac{2 i P_1(y) \int^{\pi}_{-\pi} \sin^2{\theta}\,  d\theta}{P_0(y)\int^{\pi}_{-\pi}   d\theta}=\frac{i P_1(y)}{P_0(y)}=\frac{\alpha D}{\mu f} \left(\frac{\sinh(\alpha y)}{\cosh(\alpha y)+ \zeta \cosh(\alpha L/2)} \right)
\end{equation}
From $\phi(y)$ the global order parameter 
\begin{equation}
    \psi=\int_{-L/2}^{L/2} \phi(y) \sin{\left( \frac{2 \pi y}{L}\right)} dy
\end{equation}
can then easily be determined by numerical integration.

The full steady state current vector $(J_\theta,J_y)$, which can be derived from Eq.~\ref{final} using Eq.~\ref{currydef} and $J_\theta=D_{\theta} \frac{\partial P}{\partial \theta}$, exhibits interesting structure (see Fig.~\ref{fig:higherorder}(c)). Firstly we see four points in the $y-\theta$ plane at which the average non-equilibrium current is zero. Two of them are at $y=0$ and are surrounded  by pronounced vortices of streamlines. These describe currents of particles with $\theta=\pi/2$ moving upwards (to larger $y$) and particles with $\theta=-\pi/2$ moving downwards (to smaller $y$), which are connected by currents arising from the rotation of the particle (or more precisely changes in the propulsion force orientations on the timescale of $\tau_p$). There are two other points close to the two walls where the current is zero, with the flow around these pushing particles with the minority orientation (e.g.\ pointing downwards, $\theta=-\pi/2$, for a particle that is close to the upper wall) towards the wall.

\bibliography{si_rheact}